# Separated Edge-Soliton Meditated Dynamic Switching of Vortex Chirality and Polarity

Y. M. Luo[1], Y. Z. Wu[2,3], C. Q. Yu[1], H. Li[1], J. H. Wen[1], L. Y. Zhu[1], Z. H. Qian[1] and T. J. Zhou[1*]

[1.] Center for Integrated Spintronic Devices, Hangzhou Dianzi University, Hangzhou, Zhejiang, 310018, People's Republic of China.

[2.] Department of Physics, State Key Laboratory of Surface Physics, Fudan University, Shanghai 200433, People's Republic of China.

[3.] Collaborative Innovation Center of Advanced Microstructures, Nanjing, 210093, China.

[*]Corresponding to T. J. Zhou, tjzhou@hdu.edu.cn

## Abstract

Magnetic vortices are characterized by the senses of in-plane magnetization chirality and by the polarity of the vortex core. The electrical control of vortex polarity and chirality is highly demanded not only for fundamental understanding on spin dynamics in nano-disks under different circumstances, but also for technological applications, such as magnetic non-volatile memories and spin torque oscillators for neuromorphic computing. Here we report a novel approach that enables one to electrically control both the vortex chirality and polarity with low energy consumption. Thorough micromagnetic simulations, we show that in thin nano-disks of diameter larger than 160 nm, with the presence of current-induced Oersted field, the dynamic transformation of the edge solitons is able to efficiently switch both vortex chirality and polarity with low current under certain circumstances. We then developed an approach to directly write any of the four vortex states by electrical current pulses from a random state. We further investigated the switching phase diagram as a function of disk diameters. The results show that the switching process is highly size-dependent. As disk diameter is smaller than 160 nm, the switch of VC chirality and polarity always takes place at the same time, resulting in an unchanged handedness before and after switch. Furthermore, the critical switch current can be as low as $3\times10^6 \ A/\text{cm}^2$, indicating a possible way for low current switch of vortex chirality in small disks.

# Introduction

Magnetic vortex is a flux closure domain structure in a soft magnetic nanodisk of sub-micron lateral size, which can be characterized by an in-plane curling magnetization (chirality) and a nanometer-sized vortex core (VC) with an out-of-plane magnetization (polarity) [1,2]. In general, a magnetic vortex contains four degenerated states, due to the counterclockwise (CCW, $C=1$) or clockwise (CW, $C=-1$) chirality and the upward ($P=1$) or downward ($P=-1$) polarity. The combination of the polarity and the chirality defines the handedness of the vortex with $CP=1$ and $CP=-1$ being catalyzed as left and right handed vortices, respectively. Being one of the most interesting magnetic soliton, vortex can be used as an information carrier and is currently attracting much more attention for a number of applications. Magnetic vortex have been proposed as memory bit in non-volatile storage for many years [3,4], for its multi-bit information storage and high stability [5,6]. Very recently, vortex was introduced as a building block for a robust sensor application in the automotive industry [7], where a large linear range is discovered. In addition, vortex-based spin torque nano-oscillators have been demonstrated being building blocks for neuromorphic computing, which is one of research topics towards low-power artificial intelligence application [8,9]. A full understanding on the electrical control of vortex dynamics and the switching process is a key towards those applications. In the past decades, great efforts have been made for searching effective methods to control the vortex polarity and chirality.

The VC is very stable, and a static field of above 0.5 T field is required to switch its polarity, while to switch vortex chirality requires even more energy [5,6]. Fortunately, further studied showed that it can be efficiently switched through a dynamic process [10]. When excited, the VC will be driven into gyrotropic precession, and polarity reversal happens when the VC reaches a certain critical velocity [11,12,13], through the formation and annihilation of a vortex-antivortex (VAV) pair [14,15]. Such polarity switching is energy-efficient, and an alternating (AC) field of 1.5 mT is enough to make such switch happen [10]. This efficient polarity switching has been attracted great interests, and therefore many other means, such as pulse field [16], rotating field [17], spin transfer torque (STT) [15], microwave [18] and high-frequency spin wave [19], which are able to excite spin dynamics in nano-disks, were proposed and realized. In addtion to the VAV meditated Polarity switch machanism, in 2010, Ki-Suk Lee et. al discovered a new polarity switch process, which is called edge-soliton meditated polarity switch. They found in disks with small radius, when current-excited VC reaches the disk edge, it transforms into coupled edge solitons, and VC switch happens through a fusion process [20].

On top of polarity switching, developing efficient methods to switch vortex chirality, the in-plane curling direction of disks (a 360 degree wall) is equaly important that needs to be addressed to fulfill the full potential of vortex. To switch vortex chirality by magnetic field, one needs to expel the vortex core out of the disk, and then reform a new vortex with the opposite sense of spin circulation. During the reformation, the chirality can be controlled either by exploiting an asymmetry in the structure shape [21,22,23], or by the spatial distribution of the applied magnetic field [24,25,26,27]. In addition to field driven chirality switch, electrically control the vortex chirality by spin transfer torque (STT effect) is highly desired for high density data storage. However, a very high current density is needed to switch the vortex chirality, which is on the order of $10^8 \, A/cm^2$ for switching a Py vortex with 100 nm in diameter and 6 nm in thickness [28]. Further studies show that the current can be lowered by properly utilizing the current induced Oersted field [29]. However, the strategy used in ref. 28 and 29 is not able to control the polarity at the same time. Youn-Seok Choi et al proposed a method to control both the polarity and chirality at the same time. They showed that when an out-of-plane current is applied, the STT effect and the OH field can induce polarity and chirality switch at the same time [30]. However, Youn-Seok Choi et al mainly focus on VAV meditated polarity plus chirality switch in large disks. While, high storage density demands disks with smaller size. As presented in ref. 20, as disk size scales down vortex polarity switch is meditated by edge solitons, which is faster and more energy-efficient. Nevertheless, there is no report on the chirality switch process in small disks through the formation and fusion process of edge solitons, which is the other dimension that needs to explore for high storage density. Moreover, understanding the size-dependent effect helps not only on the fundamental understanding of size-dependent spin dynamics but also for technological applications towards high storage density. Therefore, study on the chirality switch and size-effect is critically important towards technical applications.

In this manuscript, by micro-magnetic simulation, we proposed a novel edge solitons meditated vortex chirality switching process that enables the control of both vortex chirality and polarity. The VC transforms into edge solitons when it gyrates to the disk edge by out-of-plane current, and the dynamic motion, decoupling, separation and fusion of the edge solitons could induce chirality switching with the presence of current induced Oersted field. With varied current density, the dynamics of edge solitons behaves differently, and accordingly the switching process can be classified into S-state or C-state meditated switching processes. Based on this novel switching mechanism, we proposed a method to directly write any of the four degenerated vortex states by current pulses starting from a random state. We further investigated the switching phase diagram as a function of disk diameter. We show that in disks with diameters smaller than 160 nm, the C-state meditated chirality switching process is the only possible switching path, and the critical current can be as low as $3 \times 10^6 \ A/\text{cm}^2$, the same as that needed for P switch.

## Methods:

In the simulation we study the spin dynamics of the Py nanodisks. The thickness of the Py disk is fixed at $L = 12 \ nm$ if not specified, the diameter $D$ varies from $180 \ nm$ to $350 \ nm$. In such a geometry, the ground state is a vortex state. We choose vortex with CCW chirality and upward polarity ($[C, P] = [+1, +1]$) as the initial state, as shown in Fig. 1 (b). The current is applied along the z direction, we assume the current is positive (negative) when the current flow along the –z (+z) direction, and the polarization of the current is out of pane, as shown in Fig.1 (a). The out-of-plane spin polarized current can be readily generated by passing electrical current through a perpendicularly magnetized layer, such as Co/(Pd, Pt) multilayers [31], Ta/CoFeB perpendicular layer[32], or Co/Ni multilayers. The current has two effects: Firstly, it induces a circumferential Oersted field (OH) around the given current pass. Fig.1(c) shows the distribution of OH field produced by current $J = 8 \times 10^6 \ A/cm^2$; Secondly, it works as a polarized current that drives the magnetization dynamics of the Py layer due to spin transfer torque (STT effect). In our simulation, both effects are considered, and we assume the spin polarization is along –z (+z) direction when positive (negative) current is applied. We do not consider the thermo effect. At such configuration, the vortex polarity can be selectively excited: positive (negative) current excites the vortex with upward (downward) polarity because the direction of the STT depends on the relative direction of the VC and the polarization of the DC current [33].

We use OOMMF code to conduct the simulation [34], which employs the Landau–Liftshitz–Gilbert equation [35], and an additional Slonczewski spin transfer torque STT term [36]-:

$$d\vec{M}/dt = -\gamma' \vec{M} \times \vec{H}_{eff} + \alpha/|\vec{M}|[\vec{M} \times (\vec{M} \times \vec{H}_{eff})] + T_{STT} \qquad (1)$$

where $\gamma' = \gamma/(1+\alpha^2)$ with the phenomenological damping constant $\alpha$, the gyromagnetic ratio $\gamma$, and the effective field $H_{eff}$. The STT is given by $T_{STT} = (a_{STT}/\vec{M})\vec{M} \times (\vec{M} \times \hat{m}_p)$, where $a_{STT} = (1/2\pi)h\gamma PJ/(\mu_0 2eM_s L)$, $\hat{m}_p$ is the unit vector of spin polarization direction. $h$ is the Planck's constant, $J$ the current density, $\mu_0$ the vacuum permeability, $e$ the electron charge, $M_s$ the saturation magnetization, and $P$ the degree of spin polarization, and here $P = 0.7$ [3,37]. The magnetic parameters of Py is $\gamma = 2.21 \times 10^5 \, m/A \cdot s$ $\alpha = 0.01$, $M_s = 8 \times 10^5 \, A/m$, and the exchange stiffness $A = 1.3 \times 10^{-11} \, J/m$. The cell size is $2 \times 2 \times 12 \, nm^3$.

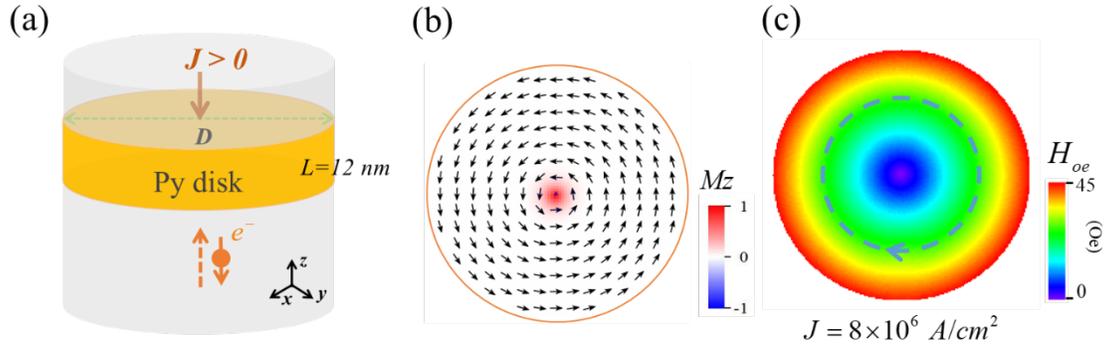

Fig.1 (a) Schematic illustration of model system. The thickness of the Py nanodisk is $L=12$ nm, and the diameter of the Py disk is $D=180$ nm if not specified. The yellow solid and doted arrow lines denote the direction of current and election flow when positive current is applied, respectively. The arrow denotes the spin polarization of the current. (b) The initial magnetic configuration. The black arrows denote the in-plane curing direction, and the color denotes the normalized out-of-plane magnetization direction, as what the color bar shows. (c) The spatial distribution of strength of OHs induced by current flow of the indicated density $J = 8 \times 10^6 \, A/cm^2$. The arrow and color-coded bar indicate the in-plane rotation sense, and the strength, respectively.

**Results and discussion**

When current is applied, VC starts to gyrate. As current reaches a critical point, edge-soliton mediated P switch takes place, which has been discussed in detail in ref. 20. With further increase of current, P-plus-C switch happens via either a C-state or an S-state, depending on the amplitude of current. Fig. 2 shows the C-state meditated P-plus-C switching process, at an excitation of positive current $J = 8 \times 10^6 \ A/cm^2$. When the current is applied, the VC is excited and gyrates around the disk center with increasing gyration speed, due to the STT effect. As soon as the VC reaches the disk edge, it transforms into a pair of coupled edge solitons. From topological point of view, the winding number ($n$), which is defined as in ref.[38], is conserved during the transformation. The winding number of VC is $+1$, while $+1/2$ for each edge soliton. Accompanied with the formation of edge solitons, the vortex also transforms into a C-state configuration, as shown in Fig. 2 (a) and (b). Such transformation has been discussed by K. Lee et al [20]. After that, the solitons move in the opposite directions along the edge, resulting in the decoupling and separating of the solitons, which is very different from what presented in ref. -20. As the solitons traveling in the opposite direction along the disk edge, the curling direction of the C-state also changes from CCW to CW, as shown in Fig. 2(c) and (d). The solitons finally meet again during the motion, leading to the merging of solutions and then transforming back into a new VC with opposite polarity and chirality ($[C,P] = [+1,+1] \rightarrow [-1,-1]$) as shown in Fig. 2(e). Upon the VC-polarity switching, the direction of STT changes from the anti-damping to the damping [31], that helps the switched VC returning to the ground state (disk center). The vortex state can be switch back by changing the current direction. Fig 2 (g) show the evolution of averaged magnetization for the in-plane $<m_x>$ and the out-of-plane component $<m_z>$ during the switch process, which shows the dynamics of VC switching process. At $t = 40 \ ns$, $<m_z>$ switched from positive to negative, indicating that the VC switch from upward to downward. $<m_x>$ oscillation show an increasing amplitude before 40 ns, but a decreasing amplitude after that, indicating that the VC gyrates to the disk edge, and then returns to the disk center after switching. Fig. 2 (h) shows the evolution of magnetic energies. There is a peak for exchange energy ($E_{ex}$), while a dip appears for dipole ($E_{de}$) energy during the switch process, indicating how the energy involved evolves during the transformation of edge solitons.

This edge-soliton-mediated polarity-plus-chirality switching has not been reported before, and it happens only at current with certain amplitude and direction. In our system, polarity switching happens when the current density is below $7\times10^6$ $A/cm^2$, or when the current direction is negative, where the edge solitons just move in the same direction and do not separate away, similar with the behavior reported in reference [20]. Such difference can be attributed to the current induced OH field. In previous study [20], the current direction is the same with OH field, while in our geometry, the OH field is opposite to the chirality. Therefore when the OH field reaches certain critical value ($7\times10^6$ $A/cm^2$, corresponding to about 15 Oe at the disk edge), the edge solitons can be separated by the OH field and then decoupled, resulting in chirality switching. As the chirality switch to the Zeeman energy favorable direction, a reduction of Zeeman energy

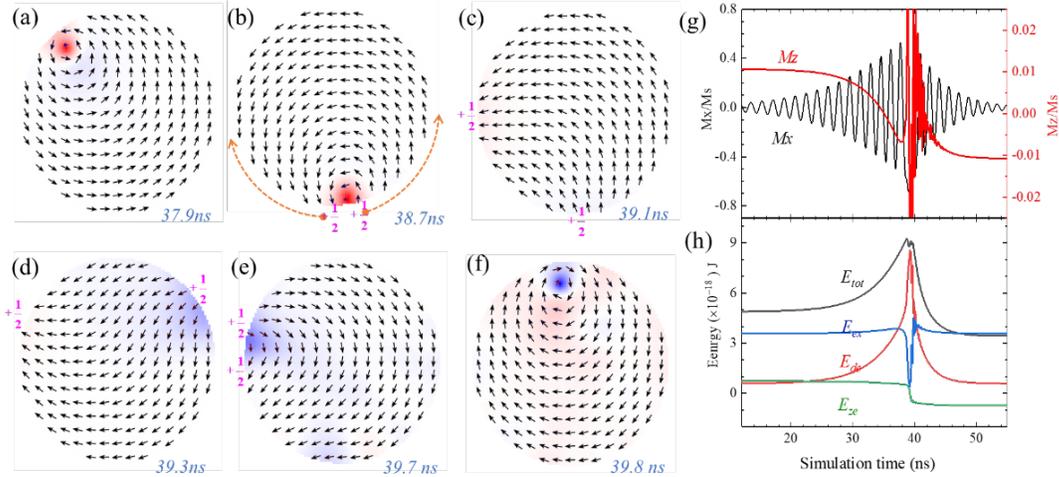

appears, and the total energy also become lower, as shown in Fig.2 (h).

Fig. 2. C-state meditated P-plus-C switch process. The current density is $J=8\times10^6$ $A/cm^2$. (a~f) Selected snapshots during the switch process. The dotted yellow arrows in (b) show the direction of the edge solitons. (g) Evolution of $<m_x>$, $<m_z>$ with time. (h) Evolution of Total ($E_{tot}$), Exchange ($E_{ex}$), Demagnetization ($E_{de}$), and Zeeman ($E_{ze}$) energy with time.

Such edge-soliton-mediated polarity-plus-chirality (P-plus-C) is energy efficient. The formation of edge solitons can greatly reduce the energy for chirality switch process.

Before the excitation of vortex, the in-plane spin direction is totally opposite (180 degree) with the OH field, the OH field induced torque ($\vec{M} \times \vec{H}$) is very small, and therefore a large field is needed. Further simulation shows that static OH field (exclude the STT effect) of about 1.25 kOe (corresponding to current density $2 \times 10^8 \, A/cm^2$) is needed to switch the chirality. However, the formation of edge solitons alters the spin structure of the disk by inducing magnetic poles at the edge, and changes the disk into C configuration. Therefore the torque ($\vec{M} \times \vec{H}$) can be greatly increased, and the critical field can be lowered by almost two orders. We show that a 45 Oe (corresponding to current density $8 \times 10^6 \, A/cm^2$) is enough to switch the chirality.

We further investigated switching process under large current density. Fig.3 shows the dynamic process when a $3 \times 10^7 \, A/cm^2$ is applied. Larger current induces stronger STT effect, and therefore the VC gyrates much faster to the disk edge and trigger the switch to happen much more quickly ($7.8 \, ns$). In addition, the stronger OH field will drive the in-plane magnetization transform from C state into S-state configuration, as shown in Fig. 3 (b) and (c). The formation of S-state also changes the number of solitons (kinks) at the disk edge, as shown in Fig. 3 (c). A new pair of edge solitons appears, just beside the original ones, which has opposite (*-1/2 and +1/2*) winding number. The solitons merge or annihilate when they move face to face along the edge. Due to the conservation of winding number, two solitons with opposite winding number (*-1/2 and +1/2*) will annihilate directly, leaving nothing but dissipating energy in the form of spin wave, while two *+1/2* solitons merge and transform into a new VC with (*+1*) winding number. According to this rule, as shown in Fig. 3 (c) and (d), although the number of solitons changes from two to six during the transformation process, the sum of winding number is conserved $\sum n = 1$. After the collision, a new VC with reversed polarity ($P = -1$) appears, and then returns back to the disk center. At the meantime, the chirality also changed from CCW to CW ($C = +1 \rightarrow C = -1$) during the motion of edge solitons, as shown in Fig. 3 (e) and (f). Fig. 3(g) and (h) show the evolution of the averaged magnetization and energies during the switch process, respectively, which display a similar behavior as what shown under lower current.

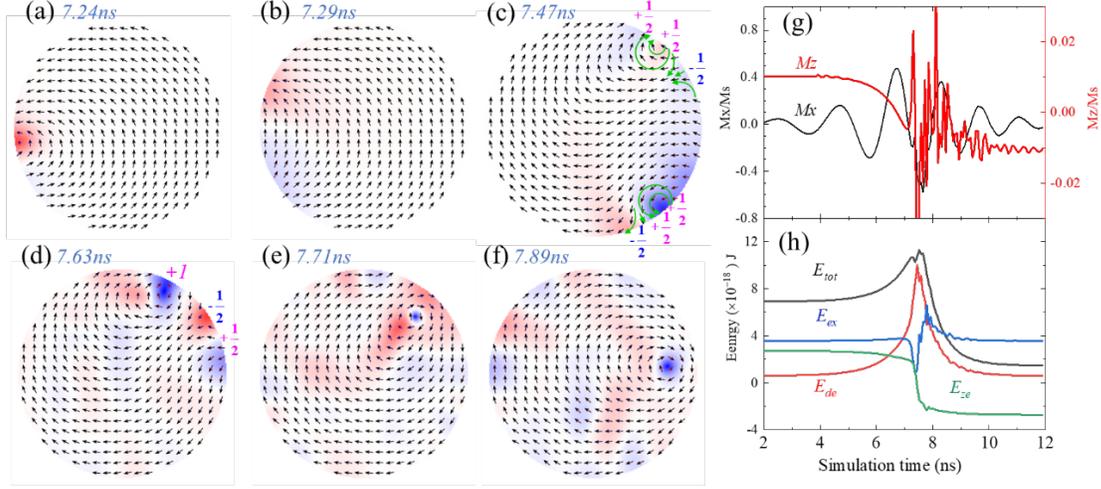

Fig. 3. S-state meditated P-plus-C switch process. The current density is $3 \times 10^7$ $A/cm^2$. (a~f) Selected snapshots during the switch process. The dotted yellow arrows in (b) show the direction of the edge solitons. (g) Evolution of $<m_x>$, $<m_z>$ with time. (h) Evolution of Total ($E_{tot}$), Exchange ($E_{ex}$), Demagnetization ($E_{de}$), and Zeeman ($E_{ze}$) energy with time.

We further investigate, systematically, the vortex switching diagram as a function of current density. Fig. 4 (a) shows the switch phase diagram of vortex state starting from upward polarity and CCW chirality ($[C,P]=[+1,+1]$) as a function of current amplitude. When the current density is lower than $J_c^I = 3 \times 10^6$ $A/cm^2$, the VC can't gyrate to the disk edge, and the vortex configuration does not change. As current increases to between $J_c^I$ and $J_c^{II} = 7 \times 10^6$ $A/cm^2$, the VC Polarity is switched, while the chirality remains unchanged. When the current goes beyond $J_c^{II}$, both the polarity and the chirality switch take place. And it should be noted that when the current is below the critical current $J_c^{III} = 2.8 \times 10^7$ $A/cm^2$, the switch is meditate by C-state, (as Fig. 2 shows); while the current excess $J_c^{III}$, the switch is meditated by S-state (as Fig.3 shows). It should be noted that in the phase diagram, the vortex is switched by positive current. On the other hand, vortex with downward polarity and CW chirality ($[C,P]=[-1,-1]$) can be switched by negative current.

Based on the novel switch mechanism discussed above, a method to directly write any of the four vortex states can be realized by applying current pulses with different amplitude and direction, as shown in Figure 4 (b). This method does not need to know the initial vortex configuration, which is highly desired for the vortex based memory application (without initialization). The polarity of VC can be controlled by using phase *II*. Due to the selective excitation of VC by out-of-plane polarized current, when a positive current with amplitude between $[J_c^I, J_c^{II}]$ is applied, the final Polarity of the VC will be +1, while a negative current between $[-J_c^{II}, -J_c^I]$ is applied, the final Polarity of the VC will be $P = -1$. Such VC switch does not change the chirality. The chirality can be controlled by using larger current. As shown in the switching phase diagram (phases III and IV of Fig. 4(a)), both the P and C can be switched. For a vortex with $P = +1$, when a positive current with amplitude larger than $J_c^{II}$ is applied, both the polarity and chirality are switched, and the final state of the VC will be $[C, P] = [-1, -1]$. On the other hand, for a vortex with $P = -1$, when a positive current with amplitude larger than $-J_c^{II}$ is applied, the final state will be $[C, P] = [+1, +1]$. According to the above analysis, the vortex states can be controlled by applying sequential current pulses with different amplitude and direction, as shown in Fig. 4(b). A vortex state $[C, P] = [+1, +1]$ can be obtained by two sequential pulses, which is denoted as ①: After applying a small positive pulse, the polarity can be initialized to $P = -1$, with chirality $C$ of either *+1* or *-1*, which can be controlled by using P-plus-C switching mode. When a large negative pulse with amplitude larger than $-J_c^{II}$ is applied, it will not only switch the polarity ($P = -1 \rightarrow P = +1$), but also switch the chirality to the OH field favorable state ($C = +1$), if it is not. The $[C, P] = [+1, +1]$ vortex state is therefore obtained. In addition, if we further apply a small positive current pulse, (denoted as ③), the polarity will be switched again ($P = +1 \rightarrow P = -1$), and $[C, P] = [+1, -1]$ state is therefore obtained. Accordingly, $[C, P] = [-1, -1]$ state can be obtained by change the current direction. Firstly a small negative is applied and then a large positive current executed (denoted as ② in Fig. 4(b)), while the $[C, P] = [-1, +1]$ state can be obtained by further switch the polarity by a small positive pulse, as shown in ④.

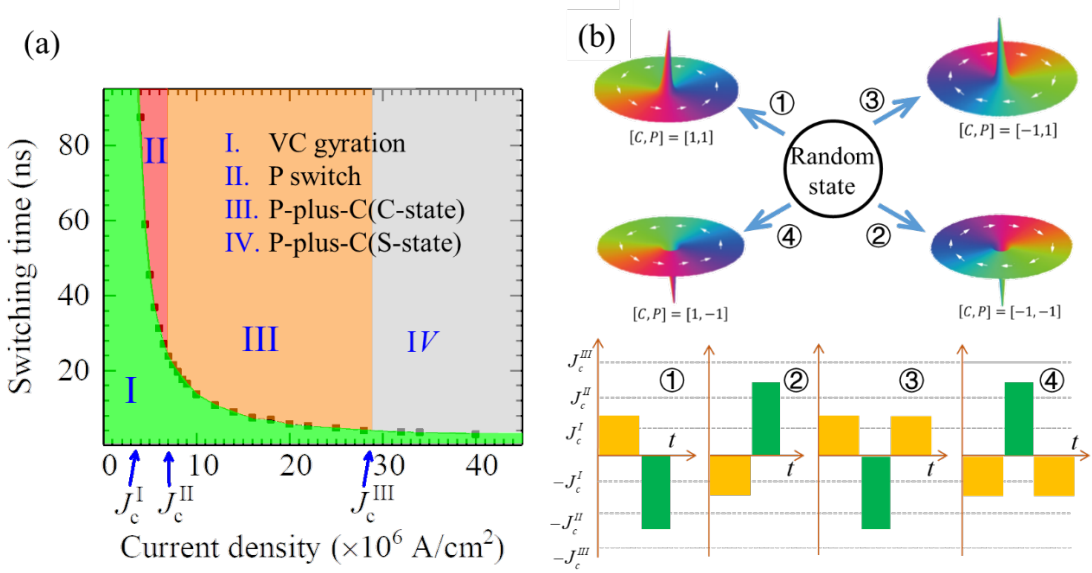

Fig. 4 (a) Vortex switching diagrams as a function of current density. The phase can be divide into four regions: I, II, III and IV, which corresponding to the region for VC gyration, P switch, C-state and C-state meditated P and C switch, respectively. $J_c^I, J_c^{II}$ and $J_c^{III}$ denote the critical currents for different switching regions. (b) Direct write the four vortex states by proper pulse current, without knowing the initial vortex configuration. The specific current pulses for writing different vortex states are denoted by ①, ②, ③ and ④, respectively, and are shown in lower part of the figure.

To investigate the scalability of such edge-solution meditated polarity and chirality switch process, we further calculate the switching diagram of disks with different diameters and thickness. Fig.5 (a) shows the switching phase as a function of the disk diameter. We choose disk with diameters between 90 nm and 350 nm. The reason is that the vortex state becomes unstable when the diameter is small than 90 nm, as what shows in the insert of Fig. 5 (a). The C state has lower energy when the diameter is below 90 nm. On the other hand, the VC switching happens inside the disk when the diameter is larger than 400 nm. The phase diagram can be divided into four regions: I, II, III and IV, corresponding to the region for VC gyration, P switch, C-state and S-state meditated P-Plus-C switch, respectively. The diagram show that when the diameter is between 160 nm and 240 nm, as the current increases, the switching behavior changes according to the following order: I→II→III→IV, which is similar with what discussed for *D=180 nm* disk. However, when the diameter is below 160 nm, there are no regions for phases II and IV.

The phase directly changes from I to III (I→III). When the diameter exceeds 240 nm, there does not exist region of phase III, and the phase transformation proceeds as I→II→IV. That is to say when the disk is smaller than 160 nm, the C-state meditated the P-plus-C switch is the only switching path, and the polarity can't be switched independently. S-state meditated switching can't be observed even when the current density reaches $4.5 \times 10^7 \, A/cm^2$. While in large disk, the S-state meditate switch dominates the P-plus-C switch process, and there is no region for the C-state meditated switch. There are mainly two facts accounting for this phenomenon: firstly, in the viewpoint of domain wall energy, in small disk, the qusi-uniform C-state is more stable than S-state. It is difficult to form S-state configuration as it has larger exchange energy, while in large disk, the situation is opposite. The S-state is more stable than C-state, and therefore the evolution of energy-favorable magnetic configuration causes the C-state (S-state) meditated switching mechanism in small (large) disks. Secondly, according to the Biot–Savart's formulation, the amplitude of OH field along the radius in the nano-pillar should exhibit a proportional behavior with respect to its distance to the pillar center. So when current is applied, large disks will have strong OH field at the edge of disks. The increase of OH field can also contribute to the S-state meditated P-plus-C switch process. Fig.5 (b) shows the switching phase diagram as a function of disk thickness. As shown in the insert of Fig. 5 (b), below 8nm, vortex becomes a metastable state. Therefore our simulation starts from 8nm thickness. When the thickness is below 10 nm, the phase transition is similar with that in small diameter disks (smaller than 160 nm), indicated by a direct phase change from I to III (I→III), without phase II. While, as the thickness is larger than 13 nm, the phase transition is similar with that in large diameter disks (larger than 240 nm), where an S-state meditated switching is dominated, and there is no range for C-state meditated switching. These behaviors can bed attribute to the energy state of the disks. As the thickness increases, S state becomes more energy favorable than C state, thus easier to form.

The coupled switching behavior in disks with diameter smaller than 160 nm or thickness below 10 nm should be emphasized for two reasons: First, previously the chirality switch is believed to be more difficult than the polarity switch, and therefore it needs larger energy. For example, in reference [30], when excited by out-of-plane current, the critical current needed for chirality switch is several times larger than the polarity switch. Previous results also showed the polarity can be switched by a small field [10], microwave [18], or spin wave [19], while the chirality can't be switched. Nevertheless, we here demonstrate that the chirality can be switched as efficiently as the polarity switch by scaling down the disk size, indicated by the same critical current, which can be, as low as $J_c^I = 3\times10^6\ A/cm^2$. Such low current induced chirality switch has unique advantage in the low energy vortex based memory application. Second, as the C-state meditated P-plus-C switch is the only possible switch path, vortex's handedness ($C\cdot P$) will be strictly unchanged after any operations, which makes it a reliable system to change vortex state while maintaining its handedness.

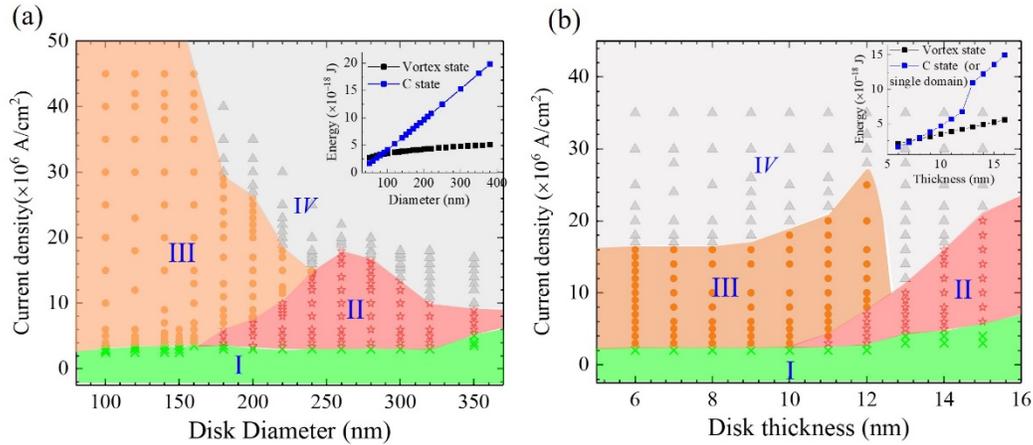

Fig. 5 (a) Vortex switching diagram as a function of disk diameter with fixed disk thickness of 12 nm. The insert shows the disk energy for Vortex state and C state as a function of disk diameter. (b) Vortex switching diagram as a function of disk thickness. The diameter is fixed at 180 nm. The insert shows the disk energy for Vortex state and C state (or single domain state) as a function of disk thickness. Both (a) and (b) can be divide into four regions: *I, II, III* and *IV*, which corresponding to VC gyration, P switch, C-state meditated P plus C switch and S-state meditated P plus C switch, respectively.

In terms of real applications, there are still several questions need to be further addressed. In the current geometry with perpendicular polarizer, the vortex chirality can't be directly read out. There are some methods which can be used to detect the chirality. For example, W. L. Lim, et.al proposed to read out the chirality by AMR effect of the Py disk, by carefully design the electrode [29]. The chirality can also be detected by adding one non-magnetic spacer and one in-plane analyzer layer (perpendicular polarizing layer/non-magnetic spacer/vortex layer/non-magnetic spacer/in-plane analyzer), which is similar with what proposed by D. Houssameddine et al [39]. The added in-plane analyzer may affect the vortex dynamics of free layer. However this can be solved by using materials of different spin polarizations and/or spin diffusion lengths for the perpendicular polarizer and planar analyzer, respectively [39].

There are concerns over the effect of intrinsic spatial dependence of spin polarization, which can be found in spin-valves and/or magnetic tunnel junction [40-42], on the vortex dynamics. We therefore carried out further simulation to study the effect of non-uniform spin polarization on the vortex dynamics through the introduction of varied non-uniformity. The simulation results show that the non-uniform spin polarization does influence the vortex dynamics through an enlarged spin transfer torque, resulting in a shorter switching time, and a lower critical current for change from C-state mediated switching to S-state meditated switching. However, the physics involved and the results obtained, including separated edge-soliton meditated chirality and polarity switching are robust and switching happens at a lower critical current due to an enlarged spin transfer torque with the introduction of non-uniform spin polarization.

## Conclusion:

In conclusion, here we reported an separated edge-soliton meditated chirality switch process in nanodisks driven by an out-of-plane currents. The dynamic transformation of the edge solitons at the disk edge plays a dominating role in such switch process. According to different dynamic behavior of edge solitons with respect to the current density, the switch process can be divided into S-state or C-state meditated switch process. Based on the novel switch mechanism, a method to directly write any of the four degenerated vortex states, starting from any random states, can be realized through applying current pulses with different direction and amplitude. The switching phase diagram as a function of disk diameter shows the scalability of such switch process. In addition, we also discover that in disks with diameters smaller than 160 nm, the chirality can be switched as efficiently as the polarity switch does, and the critical current can be as low as $J_c^I = 3 \times 10^6 \ A/\mathrm{cm}^2$. Furthermore, the vortex handedness will be strictly maintained during the switch process in small disks. This work not only provides a deeper insight into the fundamental understanding of dynamic transformations of magnetic solitons with different topological charges in nano-elements, but also offers an efficient way to control vortex chirality and polarity, which may have great application potential in vortex based spintronic devices application.

## Acknowledgments

This work was supported by the National Natural Science Foundation (Grants No. 11604066, No. 11874135, No. 61741506, No.11604132, No. 61805061) of China, and Zhejiang Science and Technology Program Project (2017C31061, 2013C31073).

J. Magn. Magn. Mater. 310, 169 (2007).